\documentclass{iopart}

\usepackage{amsfonts}
\usepackage{graphicx}
\usepackage[english]{babel}
\usepackage{subfigure}

\usepackage{microtype}

\newcommand{\be}{\begin{equation}}
\newcommand{\ee}{\end{equation}}
\newcommand{\binom}[2]{{#1 \choose #2} }

\begin{document}

\title{A Parafermionic Generalization of the Jaynes Cummings Model}

\author{Alessandro Nigro}
\ead{alessandro.nigro@mi.infn.it}
\address{Dipartimento di Fisica and I.N.F.N.\ Sezione di Milano,
Universit\`a degli Studi di Milano,
Via Celoria 16, I-20133 Milano, Italy}

\author{Marco Gherardi}
\address{Dipartimento di Fisica and I.N.F.N.\ Sezione di Milano,
Universit\`a degli Studi di Milano,
Via Celoria 16, I-20133 Milano, Italy}

\begin{abstract}
We introduce a parafermionic version of the Jaynes Cummings Hamiltonian, by coupling $k$ Fock parafermions  (nilpotent of order $F$) to a 1D harmonic oscillator, representing the interaction with a single mode of the electromagnetic field. We argue that for $k=1$ and $F\leq 3$ there is no difference between Fock parafermions and quantum spins $s=\frac{F-1}{2}$.
We also derive a semiclassical approximation of the canonical partition function of the model by assuming $\hbar$ to be small in the regime of large enough total number of excitations $n$, where the dimension of the Hilbert space of the problem becomes constant as a function of $n$. 
We observe in this case an interesting behaviour of the average of the bosonic number operator showing a single crossover between regimes with different integer values of this observable.
These features persist when we generalize the parafermionic Hamiltonian by deforming the bosonic oscillator with a generic function $\Phi(x)$; the $q-$deformed bosonic oscillator corresponds to a specific choice of the deformation function $\Phi$. In this particular case, we observe at most $k(F-1)$ crossovers in the behavior of the mean bosonic number operator, suggesting a phenomenology of superradiance similar to the $k-$atoms Jaynes Cummings model.
\end{abstract}

\maketitle

\section{Introduction}
The Jaynes Cummings (JC) model \cite{JC} realizes a description of the interaction between a quantum spin and a monochromatic mode of the electromagnetic field.
Its Hamiltonian, in the rotating-wave approximation, is (with $\hbar=1$) 
\be H= \omega a^{\dagger}a+\Delta \sigma^z+g (a \sigma^++a^{\dagger}\sigma^-)  \ee
where $a$, $a^\dagger$ are bosonic creation and destruction operators, satisfying
$[a,a^\dagger]=1$,
and $\sigma^\pm,\sigma^z$ are standard Pauli matrices, with commutation relations
\be [\sigma^+,\sigma^-]=2\sigma^z \ee
\be [\sigma^z,\sigma^\pm]=\pm \sigma^\pm \ee
It is basically obtained by coupling a single 1D harmonic oscillator to the spin matrices in such a way that the total number of excitations of the system $a^\dagger a + \sigma^z$ is conserved.
The presence of a conserved quantity allows for a finite-dimensional reduction of an otherwise infinite dimensional problem, and the model is solvable by simple methods.
The JC model can be formulated also in its multi spin version, where one has a lattice of $k$ quantum spins, with Hamiltonian
\be H= \omega a^{\dagger}a+\Delta\sum_{i=1}^k s_{i}^{z}+g \sum_{i=1}^k (a s_{i}^{+}+a^{\dagger}s_{i}^{-})  \ee
where
\be [s_{i}^{+},s_{j}^{-}]=2s_{j}^{z}\delta_{i,j} \ee
\be [s_{i}^z,s_{j}^\pm]=\pm s_{j}^\pm \delta_{i,j}  \ee
In this more general form it is equivalent to the Gaudin model \cite{Gaudin}, which can be solved by means of Bethe ansatz techniques. This model has been shown to undergo a phase transition in the thermodynamic limit between a superradiant phase, characterized by a nonvanishing expectation value of the number of ``photons'', and a normal phase where such expectation value vanishes.
It is also possible to study the higher-spin Jaynes Cummings model where one replaces the Pauli matrices by higher spin representations of the $SU(2)$ generators \cite{Gaudin}.

In this work we generalize the JC model in several directions, by replacing quantum spins with parafermions or by replacing the harmonic oscillator algebra by a deformed oscillator algebra (or both things simultaneously).
Other generalizations have been studied for instance in \cite{qdefJC,NlinJC}.
We employ this model as a simple framework for investigating the interplay between quantum oscillator algebras and parafermionic degrees of freedom.
Deformed oscillator algebras arise as generalizations of the Heisenberg-Weyl algebra \cite{Bied,Mac,Dask}.
These deformed algebras have applications for instance in nuclear and molecular physics \cite{MCD}; see \cite{deformedReview} for a review.
For the JC model, in particular, deformed oscillator algebras have been introduced in the intensity-dependent coupling formulation \cite{CEK}, thus suggesting a $q$-analog of the well-known Holstein-Primakoff transformation, mapping $SU(2)$ spins onto $q$-deformed bosonic creation and destruction operators.
The deformed oscillator algebra is generated, as in the ordinary case, by creation and destruction operators and by a number operator. However, a more general relation holds, namely $a^\dagger a=\Phi(N)$, where $\Phi$ is called \emph{deformation function}, satisfying $\Phi(0)=0$.

Parafermions have a long history and several different versions have been proposed and employed. The first instance of parafermions is due to Fateev and Zamolodchikov in the context of $\mathbb{Z}_n$-invariant 2D conformal field theory (CFT) \cite{zamo}.
These parafermions are believed to describe the scaling limit of models such as RSOS and Potts in suitable regimes.
More recently Rausch De Traubenberg studied the applications of fractional Clifford algebras to field theory, and proposed a CFT possessing fractional supersymmetry \cite{traub1,traub2}.
In \cite{traub1} both the fractional clifford algebra (with idempotent generators) and the para-Grassmann algebra (with nilpotent generators, thus generalizing fermions) were defined.
More recently, parafermions became again object of interest as a tool for studying lattice models   such as $\mathbb{Z}_n$-invariant spin chains \cite{fendley1,fendley2}, Potts models, or even the more general $\tau_2$ model \cite{bax,perk}. It is not difficult to see that the definition of fractional Clifford algebra given by Rausch De Traubenberg indeed coincides with that given by Fendley in \cite{fendley1}, although the latter employs a Jordan-Wigner transformation in its definition. 
These recent works all employ
\emph{idempotent} parafermions, which loosely speaking are generalization of Majorana fermions. To our knowledge, the relation of these parafermions defined on a lattice with the parafermions of Fateev and Zamolodchikov has not been clarified.
%%%%%%%%%
It is also worth mentioning that there exist several physical situations and applications in which other particles (or also effective degrees of freedom) called \emph{anyons} appear;
they are not necessarily parafermions, but they have nontrivial exchange statistics too.
A typical setting in which anyons appear is the fractional quantum Hall effect,
with applications in
topological quantum computing, a field of great technological impact
(the literature at this ``mesoscopic'' level is huge; see for instance \cite{meso1,meso2,meso3,meso4,meso6}).
%%%%%%%%%%%

In this work we will focus on \emph{nilpotent} parafermions.
It is easy to see that by dressing a set of commuting deformed oscillator modes $a_i$ 
with a generalized Jordan-Wigner transformation
\be \theta_j=\prod_{i<j}q^{N_i}a_j \ee
where $N_i$ is the number operator
\be N_i=\Phi^{(-1)}(a_i^\dagger a_i)  \ee
one obtains a set of creation and destruction operators that $q-$commute.
To obtain (para) fermions one needs an additional algebraic structure, that is the (generalized) Clifford algebra,
identified by an integer $F$:
\be 
\sigma_i^F=1
\ee
\be  \sigma_i\sigma_j=e^{\frac{2\pi i}{F}}   \sigma_j\sigma_i ,\quad i<j \ee
The generalized $\sigma$ matrices are defined as:
\be \sigma_1= \left(
%\begin{matrix} 
\begin{array}{ccccc}
0 & 1 &0 &\ldots &0  \\
0&  0 &1  &\ldots &0 \\  
\vdots&  &\ddots  &\ddots &\vdots \\  
0&  0 &\ldots  &0 &1 \\                                          
1& 0 & 0&\ldots& 0 
\end{array} 
%\end{matrix} 
\right)  \ee
                                             
\be \sigma_3=\left(
%\begin{matrix} 
\begin{array}{ccccc}
1 &0 &\ldots &0  \\
0&  e^{\frac{2\pi i}{F}} &\ldots   &0\\   
\vdots&   & \ddots &  \\                                       
0&  0 & \ldots&e^{\frac{2\pi i (F-1)}{F}} 
\end{array}
%\end{matrix} 
\right)  \ee            
                                             
\be  \sigma_2= \sigma_3\sigma_1  \ee
By identifying the $q$ in the Jordan-Wigner with $e^{\frac{2\pi i}{F}}$ and recognizing that
\be \sigma_3=q^{N} \ee
one can show that the deformed algebra,
%%%%%%%%%%%%
subject to the condition $\Phi(F)=0$,
%%%%%%%%%%%%
is obtained from the matrices of the generalized Clifford algebra through
\be  a=\frac{\sqrt{\Phi(N+1)}}{1-q^{N+1}}(\sigma_1-q\sigma_2) \ee
In this formula, it is understood
that when the denominator is vanishing, one should take the limit for $q$ going to a root of unity. Since $\Phi(F)=0$ one then has $a^F=0$ and the variables $\theta_i$ can be now called parafermions; algebras of this kind appear for example in \cite{OrtizPF} and \cite{exclstat}.
 In particular we shall consider in this work only Fock parafermions (introduced in \cite{OrtizPF}), which correspond to a limit in which the deformation function $\Phi$ is not invertible. In what follows we shall avoid to employ the Jordan-Wigner transformation, and we will define the Fock parafermions directly by their action on a basis of the Hilbert space.

The outline of the work is as follows. We first introduce Fock parafermions and their properties, then we define the parafermionic JC model with undeformed harmonic oscillator, and present the solution in the case $F=2$ (and generic $k$), and in the case $F=3$, $k=1$ (we argue that also $F=4$ is exactly solvable when $k=1$, although we do not give the explicit form of the solution). We consider the further generalization whereby we deform the harmonic oscillator, and find that the model again can be analytically solved in the cases above.
Then, by inserting Planck's constant $\hbar$ into the formulas, we show that a semiclassical expansion to the canonical partition function of the model can be obtained, that turns out to be a surprisingly good approximation, even in the regime where $\hbar$ is not small. The accord is checked by means of numerical computations, through direct diagonalization of the Hamiltonian.
These approximants of the partition function of the model are derived in the regime of large enough total number of excitations $n$ of the system, in which the dimension of the Hilbert space of the problem saturates  as a function of $n$.
The average of the number operator associated to the harmonic oscillator is shown numerically to exhibit a smooth crossover between a superradiant regime and a normal regime.
Finally, we show that a richer structure of crossovers opens up when the $q$-deformation of the harmonic oscillator is turned on, revealing up to $k(F-1)+1$ different regimes in the limit where the dimension of the Hilbert space saturates.

\section{Fock parafermions and the parafermionic JC model}

As mentioned in the introduction, different kinds of parafermionic algebras have been studied. Here we concentrate on the so-called Fock parafermions, defined through an algebra of nilpotent variables, thus generalizing the Grassmann algebra:
\be  \label{eq:q_commutation}
\theta_i \theta_j=q  \theta_j \theta_i, \quad i<j  
\ee
\be \label{eq:nilpotent_F}
( \theta_i)^{F}=0  
\ee
\be q=e^{\frac{2\pi i}{F}}  \ee
We will suppose that there exists a vacuum state $\big|0\big>$ such that:
\be   \theta_i\big|0\big>=0  \ee
By acting on the vacuum state with the creation operators --- at most $F-1$ times for each creation operator, because of the constraint imposed by (\ref{eq:nilpotent_F}) --- one can define states $\big|j\big>$ for the single site problem in the following way (we shall generalize this notation to many sites shortly):
\be( \theta^{\dagger})^{j} \big|0\big>= \big|j\big> \ee
These rules can be used to compute matrix elements involving creation and destruction operators.
The generic state in the Fock space is  labelled by quantum numbers $\mathcal{P}=(i_1,\ldots,i_k)$, where each $i_m$ can take values from $0$ to $F-1$; of course they have the desired multiplicity $F^k$. In terms of these quantum numbers --- which are simply the occupation numbers of the different allowed creation operators --- it is possible to write the generic state created by a monomial in the creation operators of the Fock space as: 
\be \Big|\mathcal{P}\Big>=(\theta_k^{i_k})^{\dagger}\ldots(\theta_1^{i_1})^{\dagger}\Big|0\Big>  \ee
In this basis it is possible to reconstruct the matrix element of the destruction operators between states labelled by different occupation numbers  $\mathcal{P}$ in the following way:
\be   \Big<\mathcal{P}\Big|\theta_m\Big|\mathcal{P}' \Big>=q^{-\sum_{s=m+1}^{k}i_s}\delta_{\mathcal{P}\cup\{m\},\mathcal{P}'}  \ee
where $\mathcal{P}\cup\{m\}$ denotes the quantum number $\mathcal{P}$ with its $m-$th entry increased by one.
The exponential factor in the matrix element comes from the $q-$commutation relations (\ref{eq:q_commutation}) needed to move the operator $\theta_m$ toward the left until it reaches the standard-ordered position inside the monomial labelled by $\mathcal{P}$. Remark that the exponential factors are crucial for obtaining $q-$commuting generators $\theta_m$, whereas the Kronecker delta between different quantum numbers ensures separately only the nilpotency of each $\theta_m$.

It is possible to build number operators of the form:
\be N_i=\sum_{s=1}^{F-1}(\theta^{\dagger}_i)^s\theta^s_i  \ee
satisfying the usual commutation relations:
\be  [N_i,\theta_j]=-\delta_{i,j}\theta_j   \ee
\be  [N_i,\theta^{\dagger}_j]=\delta_{i,j}\theta^{\dagger}_j   \ee
We will also define a total number operator $\mathcal{N}$:
\be  \mathcal{N}=\sum_{i=1}^k N_i   \ee
with eigenvalues $W(\mathcal{P})$ satisfying
\be   \mathcal{N}\Big|\mathcal{P} \Big>=W(\mathcal{P})\Big|\mathcal{P} \Big>  \ee
\be W(\mathcal{P})=\sum_{m=1}^{k}i_m  \ee
Sometimes the notation $\mathcal{N}(\theta,\theta^{\dagger})$ will be used to highlight the explicit dependence on the parafermionic operators.

We are going to couple $k$ Fock parafermions to one Harmonic oscillator by means of the following self adjoint Hamiltonian:
\be \label{eq:parafermionic_JC}
H= \omega a^{\dagger}a+\Delta\mathcal{N}(\theta,\theta^\dagger)+g \sum_{i=1}^k (a \theta_{i}^{\dagger}+a^{\dagger}\theta_{i})  
\ee
We remark that the coupling between parafermions and the bosonic creation and destruction operators 
in (\ref{eq:parafermionic_JC}) is linear, in contrast to the usual fermionic case, whereby the electromagnetic
field couples to fermionic bilinears.
A similar linear Hamiltonian --- with $F=2$ --- has been proposed in \cite{linferm} as an approximate
description of the Jaynes-Cummings-Hubbard model, accounting for essentially the same phenomenology
as the bilinear form.
%%%%%%%%%%%%%%%
The fermionic approximation in \cite{linferm} is especially successful in that it is exactly solvable by Fourier transform. 
However, this is not the case for the general parafermionic case, 
where the Hamiltonian is decomposed as a sum of terms that do not commute 
and hence the solution is not straightforward. 
By ``approximation'' we refer to the fact the the parafermionic JC model describes the same phenomenology
as the ordinary JC model
(we have checked through simulations that the same behaviour exposed in the following sections is observed).
It is also worth pointing out that performing an inverse Jordan Wigner transformation on the parafermionic JC model
brings to a description in terms of bosonic degrees of freedom. 
The bosonic model then contains
complicated many-body interactions brought about by the presence of the string factor,
which does not cancel out in the interaction term.
%%%%%%%%%%%%%%%
Due to the $q-$commuting nature of Fock parafermions, the Hamiltonian 
(\ref{eq:parafermionic_JC}) is not to our knowledge solvable by Bethe ansatz.
Therefore we shall attempt to generalize the simple solution method for the 1-atom JC model to this case.
It will turn out that the model with $k=1$ is analytically solvable for the first few values of $F$. Actually we observe that for $k=1$ and $F=2,3$ there is no difference between Fock parafermions and quantum spins of spin $j=1/2,1$ respectively. 
This is a consequence of the fact that the creation and annihilation operators for parafermions and spins in these very special cases are represented by the same matrices.

\section{Exact spectrum for special cases}

Let $N_{tot}$ denote the total number of excitations:
\be  N_{tot}=a^\dagger a+\mathcal{N}(\theta,\theta^\dagger)  \ee
It is not difficult to see that this operator commutes with the $k$-atom parafermionic JC Hamiltonian. 
As a consequence of this conservation, the eigenvectors of $N_{tot}$ can be used to build a matrix representation of the Hamiltonian. They take the following form:
\be \Big|n,\mathcal{P}\Big>=\Big|n-W(\mathcal{P})\Big>\otimes\Big|\mathcal{P}\Big>  \ee
where only partitions $\mathcal{P}$ with weight $W(\mathcal{P})\leq n$ are kept as a basis of the fixed particle-number Hilbert  space.

The dimensions of the eigenspaces with
\be N_{tot} \Big|n,\mathcal{P}\Big>=n \Big|n,\mathcal{P}\Big>  \ee
can be easily computed by the following generating function:
\be \label{eq:generating_size}
J(x)=\textrm{Tr}x^{N_{tot}}=\frac{(\sum_{s=0}^{F-1}x^s)^k}{(1-x)}=  \frac{(1-x^F)^k}{(1-x)^{k+1}} 
\ee
and they are generated by expanding the function $J(x)$:
\be  J(x)=\sum_{n=0}^\infty d_n(k)x^n  \ee
or more explicitly:
\be d_n(k)= \sum_{s=0}^{\textrm{min}(k,\frac{n}{F})}(-1)^{n-s(F-1)}\binom{k}{s}\binom{-k-1}{n-s F}\ee
For example, for $F=4$ and $k=3$ one has:
\be 
%\begin{split}
%\begin{eqnarray*}
\eqalign{
J(x)=&1+ 4x+10x^2+20x^3+32x^4+44x^5\\
&+54x^6+60x^7+63x^8+64x^9+64x^{10}+\ldots  
}
%\end{eqnarray*}
%\end{split}
\ee
Due to the ortogonality relations
\be\Big<n,\mathcal{P}' \Big|m,\mathcal{P}\Big>=\delta_{n,m}\delta_{\mathcal{P},\mathcal{P}'} \ee
one obtains the following matrix representation of the Hamiltonian labelled by the quantum number $n$
%\be\begin{split} &\Big<n,\mathcal{P}' \Big|H\Big|n,\mathcal{P}\Big>=(\omega(n-W(\mathcal{P}))+\Delta W(\mathcal{P}))\delta_{\mathcal{P},\mathcal{P}'} +\\&+g \sum_{l=1}^k \Big(\sqrt{n+1-W(\mathcal{P})}e^{-\frac{2\pi i}{F}\sum_{s=l+1}^{k}i_s}\delta_{\mathcal{P}' \cup\{l\},\mathcal{P}}+\\&+\sqrt{n-W(\mathcal{P})}e^{\frac{2\pi i}{F}\sum_{s=l+1}^{k}i'_s}\delta_{\mathcal{P}\cup\{l\},\mathcal{P}'}\Big)\end{split}\ee
\be
%\begin{eqnarray}
\eqalign{
\fl 
\Big<n,\mathcal{P}' \Big|H\Big|n,\mathcal{P}\Big>=&\left[\omega\left(n-W(\mathcal{P})\right)+\Delta W(\mathcal{P})\right]\delta_{\mathcal{P},\mathcal{P}'} +\\
&g \sum_{l=1}^k \Big(\sqrt{n+1-W(\mathcal{P})}e^{-\frac{2\pi i}{F}\sum_{s=l+1}^{k}i_s}\delta_{\mathcal{P}' \cup\{l\},\mathcal{P}}\\
&\phantom{g \sum_{l=1}^k \Big(}+\sqrt{n-W(\mathcal{P})}e^{\frac{2\pi i}{F}\sum_{s=l+1}^{k}i'_s}\delta_{\mathcal{P}\cup\{l\},\mathcal{P}'}\Big)
}
%\end{eqnarray}
%\end{split}
\ee
Now by choosing a basis for the states $\mathcal{P}$, we can represent the $k$-atoms parafermionic JC Hamiltonian within the $n-$th energy level by a matrix $H_n$ whose size is generated by $J(x)$.
By expression (\ref{eq:generating_size}), it can be seen that this size saturates to $F^k\times F^k$ when $n>k(F-1)$.
This has a useful consequence: each $H_n$ can then
be diagonalized for $k=1$ by analitical means up to $F=4$ and all the energy levels can be obtained exactly, also in the regime $n\leq F-1$. 

We shall give the explicit solution for $F=2,3$ in the regime $n>F-1$.
The case $F=4$ requires solving algebraic equations of degree $4$,
so the solutions are complicated, and not illuminating, so we will not write them down explicitly.
For $F>4$ or whenever $k>1$, these matrices have to be diagonalized numerically; we will get to this problem in the next section.
For $k=1$ and $F=3$ the expressions are already rather involved, but they can be slightly simplified by defining the following quantity:
\begin{equation}
%\begin{split}
\fl 
\Omega_n(\omega,\Delta,g)=
\Big[\sqrt{-108(g^2(2n-1)+(\Delta-\omega)^2)^3+729 g^4(\Delta-\omega)^2}
+27g^2(\omega-\Delta)\Big]^{\frac{1}{3}}
%\end{split}
\end{equation}
The energy levels then can be written as
\be
%\begin{split}
\fl
E_n^{(l)}=\Delta+(n-1)\omega+ e^{-\frac{2\pi i}{3}(l)}\frac{\Omega_n}{3 \ 2^{\frac{1}{3}}}
+e^{\frac{2\pi i}{3}l}\frac{2^{\frac{1}{3}}(g^2(2n-1)+(\Delta-\omega)^2)}{\Omega_n}
%\end{split}
\ee
with $l=0,1,2$.
For $F=2$ and generic $k$ the solution can be obtained by Fourier transform.
In fact, since the Fourier transform of the interaction term is exactly the zero mode of the Fourier transform of the Fermi modes, the Hamiltonian becomes that of a $k=1$ JC model plus some number operators, so that the solution is straightforward:
\be 
%\begin{split}
\fl 
E^\pm_{n,l}=\frac{1}{2}\Bigg((2l+1)\Delta+(2(n-l)-1)\omega
\pm\sqrt{4kg^2(n-l)+(\Delta-\omega)^2}\Bigg)
%\end{split}
\ee
where $l=0,\ldots,k-1$, and
each $E^\pm_{n,l}$ has a degeneracy that exactly equals $\binom{k-1}{l}$. 
As this solution is essentially the same as the JC model plus some integers, the physics of this Hamiltonian is expected to be basically the same as the JC model.

\section{Coupling parafermions to a deformed bosonic oscillator\label{section:deformed}}

In this section we explore the coupling of Fock parafermions to a $q$-deformed oscillator, by using again the generalized JC Hamiltonian (\ref{eq:parafermionic_JC}).
This case presents a richer behavior than the undeformed one.

The deformed bosonic oscillator is defined quite generally by assigning a deformation (or structure) function $\Phi(x)$, in terms of which the algebra of creation-destruction operators $a,a^\dagger$ and the number operator $N$ has the following properties:
\be [N,a]=-a  \ee 
\be [N,a^\dagger]=a^\dagger  \ee
\be  a^\dagger a=\Phi(N)  \ee 
\be  a a^\dagger=\Phi(N+1)  \ee
A commonly-found simple form for the structure function is
\be
\Phi(x) =[x]_q=\frac{q^{-x}-q^x}{q^{-1}-q},
\ee
i.e, the so-called \emph{$q$-number}.
The action of operators on eigenstates of the number operator is given by the following relations:
\be N\Big|n\Big>=n\Big|n\Big>   \ee
\be  a\Big|n\Big>=\sqrt{\Phi(n)}\Big|n-1\Big> \ee 
\be  a^\dagger\Big|n\Big>=\sqrt{\Phi(n+1)}\Big|n+1\Big> \ee
Notice that in order for the vacuum to be annihilated by the destruction operator one must obviously require $\Phi(0)=0$.\\
We generalize the parafermionic JC Hamiltonian by replacing the undeformed $a$ and $a^\dagger$ with their deformed version described above. 
Inclusion of a more general kinetic term for the deformed oscillator
(a generic function of $a^\dagger a$ plus another generic function of $\mathcal{N}$)
is not expected to change the relevant features of this generalization.
The total number operator $N_{tot}$ commutes with the Hamiltonian as before; 
thus a matrix representation of the Hamiltonian can be built within the eigenspaces with fixed eigenvalue of the total number, on the same basis as before:
\be
%\begin{split}
\eqalign{
\Big|n,\mathcal{P}\Big>=\Big|n-W(\mathcal{P})\Big>\otimes\Big|\mathcal{P}\Big>\\
\Big<n,\mathcal{P}' \Big|m,\mathcal{P}\Big>=\delta_{n,m}\delta_{\mathcal{P},\mathcal{P}'}
}
%\end{split}
\ee
%we then have:
%\be\Big<n,\mathcal{P}' \Big|m,\mathcal{P}\Big>=\delta_{n,m}\delta_{\mathcal{P},\mathcal{P}'} . \ee
One obtains the following matrix representation of the Hamiltonian, labelled by the quantum number $n$:
\begin{equation}
%\begin{split} 
\eqalign{
\fl
\Big<n,\mathcal{P}' \Big|H\Big|n,\mathcal{P}\Big>=&\left[\omega\Phi(n-W(\mathcal{P}))+\Delta W(\mathcal{P})\right]\delta_{\mathcal{P},\mathcal{P}'}\\
&+g \sum_{l=1}^k \Big(\sqrt{\Phi(n+1-W(\mathcal{P}))}e^{-\frac{2\pi i}{F}\sum_{s=l+1}^{k}i_s}\delta_{\mathcal{P}' \cup\{l\},\mathcal{P}}\\
&\phantom{+g\sum\Big(}+\sqrt{\Phi(n-W(\mathcal{P}))}e^{\frac{2\pi i}{F}\sum_{s=l+1}^{k}i'_s}\delta_{\mathcal{P}\cup\{l\},\mathcal{P}'}\Big)
}
%\end{split}
\end{equation}
The structure of the reduced Hamiltonian does not change with respect to the undeformed case;
the deformation function $\Phi$ enters in a simple way into the expression.
It turns out that the model for generic $k$ and $F=2$ is still exactly solvable.
In the saturated regime where $n\geq k(F-1)\equiv k$, its eigenvalues are given by
\be
%\begin{split}
\fl
E^\pm_{n,l}=\frac{1}{2}\Bigg((2(k-l)-1)\Delta\pm R_{n,l}(\omega,\Delta,g)
+\left[\Phi(n-k+l)+\Phi(n-k+l+1)\right]\omega\Bigg)
%\end{split}
\ee
for $l=0,\ldots,k-1$, with
\begin{equation}
%\begin{split}
\eqalign{
\fl
R_{n,l}(\omega,\Delta,g)=\Big[\left(\Delta+\omega\Phi(n-k+l)\right)^2
+\Phi(n-k+l+1)\\
\cdot\Big(4kg^2-2\omega\Delta-2\omega^2\Phi(n-k+l)
%\phantom{+\Phi(n-k+l+1)\Big(}
+\omega^2\Phi(n-k+l+1)\Big)\Big]^{1/2}
}
%\end{split}
\end{equation}
Again, each eigenvalue has degeneracy $\binom{k-1}{l}$.
It is still also possible to solve for the spectrum of the $k=1$, $F=3$ Hamiltonian; its eigenvalues can be found but their form is rather complicated and therefore we shall omit it.

%%%%%%%%%%%%%%%
We end this section by a simple observation connecting the deformed
oscillator algebra with higher spin representations of $SU(2)$.
It will enable the numerical study of the ordinary higher-spin JC model in a quick
way by use of the same techniques employed for the parafermionic and deformed JC
(see Sec.~\ref{section:numerical}).
We could not find this connection described in the literature,
so it is worth stating it explicitly.
The matrices $\sigma^+, \sigma^-, \sigma^z$ in their higher spin-$j$ representations can be realized 
in terms of elements of the deformed algebra with the following special deformation function:
\be \Phi(x)=x(F-x)  \ee
This is the structure of what is called the \emph{parafermionic oscillator algebra} \cite{deformedReview}.
Notice that $a^F=0$, since $\Phi(F)=0$.
If we now identify $j=\frac{F-1}{2}$, the higher spin matrices are indeed given by:
\be  \sigma^+=a^\dagger \ee
\be  \sigma^-=a \ee
\be \sigma^z=N-\frac{F-1}{2}\,\mathbb{I}  \ee
%%%%%%%%%
The foregoing observations implies that the higher spin JC Hamiltonian can be built in the eigenspace with fixed total number of excitations $n$ by means of the following matrix element:
\begin{equation}
%\begin{split} 
\eqalign{
\fl
\Big<n,\mathcal{P}' \Big|H\Big|n,\mathcal{P}\Big>=\left[\omega\Phi(n-W(\mathcal{P}))+\Delta W(\mathcal{P})\right]\delta_{\mathcal{P},\mathcal{P}'}\\
+g \sum_{l=1}^k \Big(\sqrt{\Phi(n+1-W(\mathcal{P}))}\sqrt{i_l(F-i_l)}\delta_{\mathcal{P}' \cup\{l\},\mathcal{P}}\\
+\sqrt{\Phi(n-W(\mathcal{P}))}\sqrt{(i_l+1)(F-1-i_l)}\delta_{\mathcal{P}\cup\{l\},\mathcal{P}'}\Big)
}
%\end{split}
\end{equation}
We have used this shortcut for comparing the predictions of the parafermionic and the higher-spin models;
results are briefly stated in Sec.~\ref{section:numerical}.
%%%%%%%%%%%%%%%

\section{Semiclassical approximation}

In this section we are going to discuss the partition function of this parafermionic Jaynes-Cummings model. Due to the nontrivial form of the eigenvalues of the Hamiltonian, the hope to find an analytic closed form for the partition function is rather small. 
However, recall that in all calculations so far we set $\hbar=1$.
The idea behind the semiclassical approximation is rather to view $\hbar$ as a small parameter and to consider only up to the linear terms in the expansion of the energy levels.
Planck's constant is reinstated by choosing a dimensionful deformation function for the boson:
\be \Phi(x)=\hbar x  \ee
It is then possible to recognize, by directly writing down the expansion, 
that the eigenvalues for $F=2$ in the regime $n>k$ scale as:
\begin{equation}
%\begin{split}
\fl
E^{s}_{n,l}\sim \frac{1}{2\Delta}\Big[
2 g^2 k s \hbar  (l+n-k+1)+\Delta ^2 (2 k-2 l+s-1)
+\Delta  \omega  \hbar  (2 l+2 n-2k-s+1)\Big]
%\end{split}
\end{equation}
with $s=\pm 1$ and $l=0,\ldots,k-1$. 
For each $l$ the degeneracy is $\binom{k-1}{l}$, so for $F=2$ and generic $k$ one has:
\begin{equation}
Z(2,k,\hbar,\omega,\Delta,g)=\sum_{l=0}^{k-1}\binom{k-1}{l}\left(e^{- E^{+}_{n,l}}+e^{- E^{-}_{n,l}}\right)
\end{equation}
which can be explicitly evaluated to give:
%\begin{widetext}
\begin{equation}
\label{eq:semiclassical_k}
%\begin{split}
\eqalign{
\fl
Z(2,k,\hbar,\omega,\Delta,g)=&\:
\frac{\left(e^{\Delta -\frac{g^2 k \hbar }{\Delta }-\omega  \hbar }+1\right)^k}{e^{\Delta }+e^{\hbar  \left(\frac{g^2
   k}{\Delta }+\omega \right)}}
   e^{\frac{\hbar  \left(\Delta  \omega +(k-n) \left(\Delta  \omega +g^2
   k\right)\right)}{\Delta }-\Delta  k}\\
&+\frac{\left(e^{\Delta +\frac{g^2 k \hbar }{\Delta
   }-\omega  \hbar }+1\right)^k}{e^{\Delta +\frac{g^2 k \hbar }{\Delta
   }}+e^{\omega  \hbar }}
   e^{\Delta +\frac{g^2 k \hbar  (-k+n+1)}{\Delta}
   -\Delta  k+\omega  \hbar  (k-n)}
%\phantom{+\frac{\left(e^{\Delta +\frac{g^2 k \hbar }{\Delta
 %  }-\omega  \hbar }+1\right)^k}{e^{\Delta +\frac{g^2 k \hbar }{\Delta
  % }}+e^{\omega  \hbar }}
  % \exp \Bigg(}
}
%\end{split}
\end{equation}
%\end{widetext}
For generic $F$ and $k=1$ instead we obtain:
\begin{equation}
\label{eq:semiclassical_F}
%\begin{split}
\eqalign{
\fl
Z(F,1,\hbar,\omega,\Delta,g)=&\:
e^{-\frac{\hbar n \left(\Delta  \omega -g^2\right)}{\Delta }}+e^{\Delta  (1-F)-\frac{g^2 (n-F+2) \hbar }{\Delta }-(n+1-F)
   \omega  \hbar }\\
&+\sum_{s=1}^{F-2}e^{-\omega  \hbar  (n-s)-\frac{g^2 \hbar }{\Delta }-\Delta  s}
}
%\end{split}
\end{equation}
We will check the accuracy of our approximation in the next section, by comparing the predictions (\ref{eq:semiclassical_k}) and (\ref{eq:semiclassical_F}) with numerical computations.
The partition functions obtained in this small-$\hbar$ approximation show a crossover between 2 regimes,
approximately at $\omega=\frac{\Delta}{\hbar}$.
The behavior for generic $F$ and $k$, instead, is more complicated and is not captured by the semiclassical phenomenology.

\begin{figure}
\centering
\begin{subfigure}
  \centering
  \includegraphics[width=0.6\linewidth]{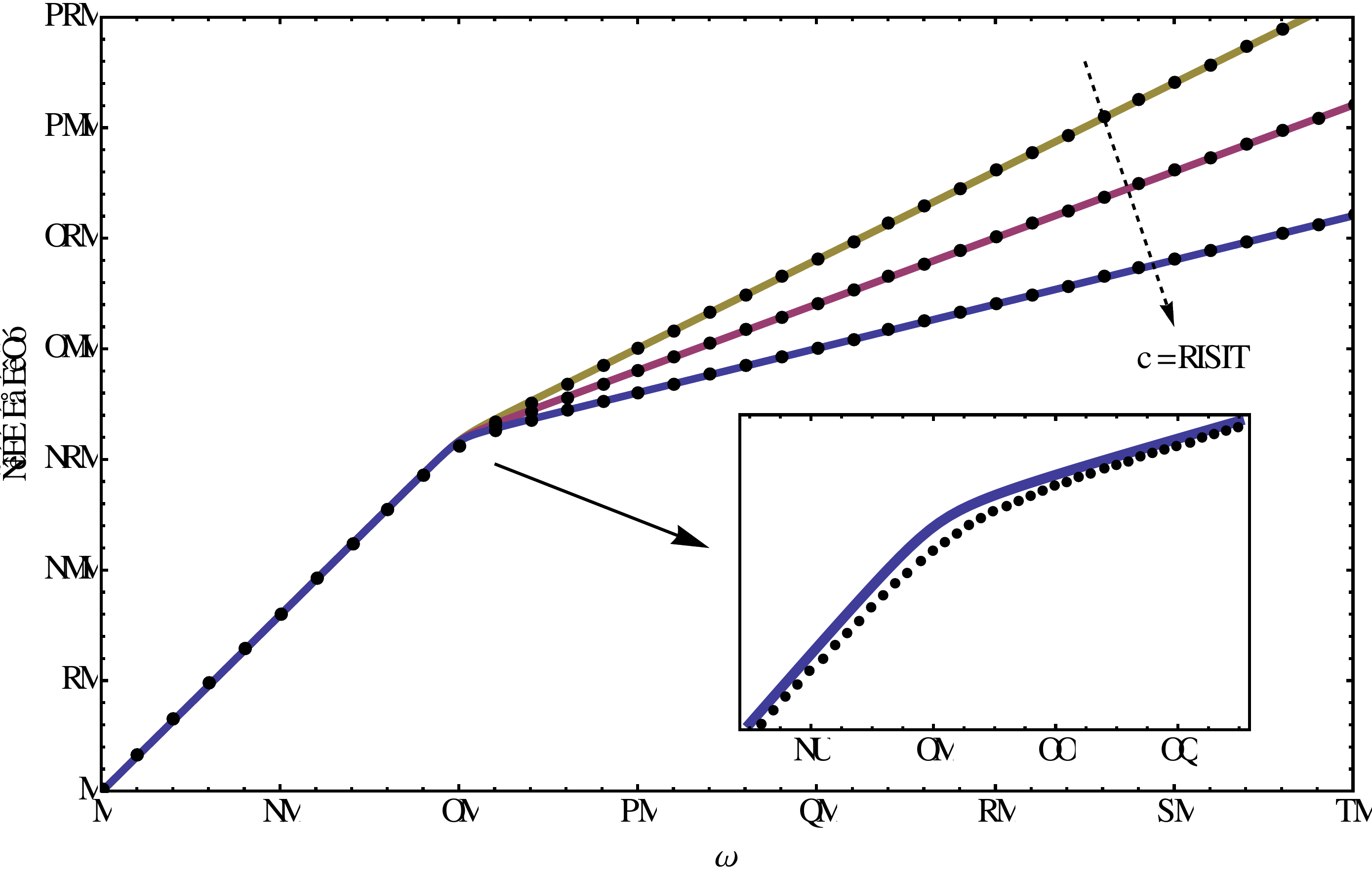}
\end{subfigure}%
\begin{subfigure}
  \centering\hspace{1em}
  \includegraphics[width=0.6\linewidth]{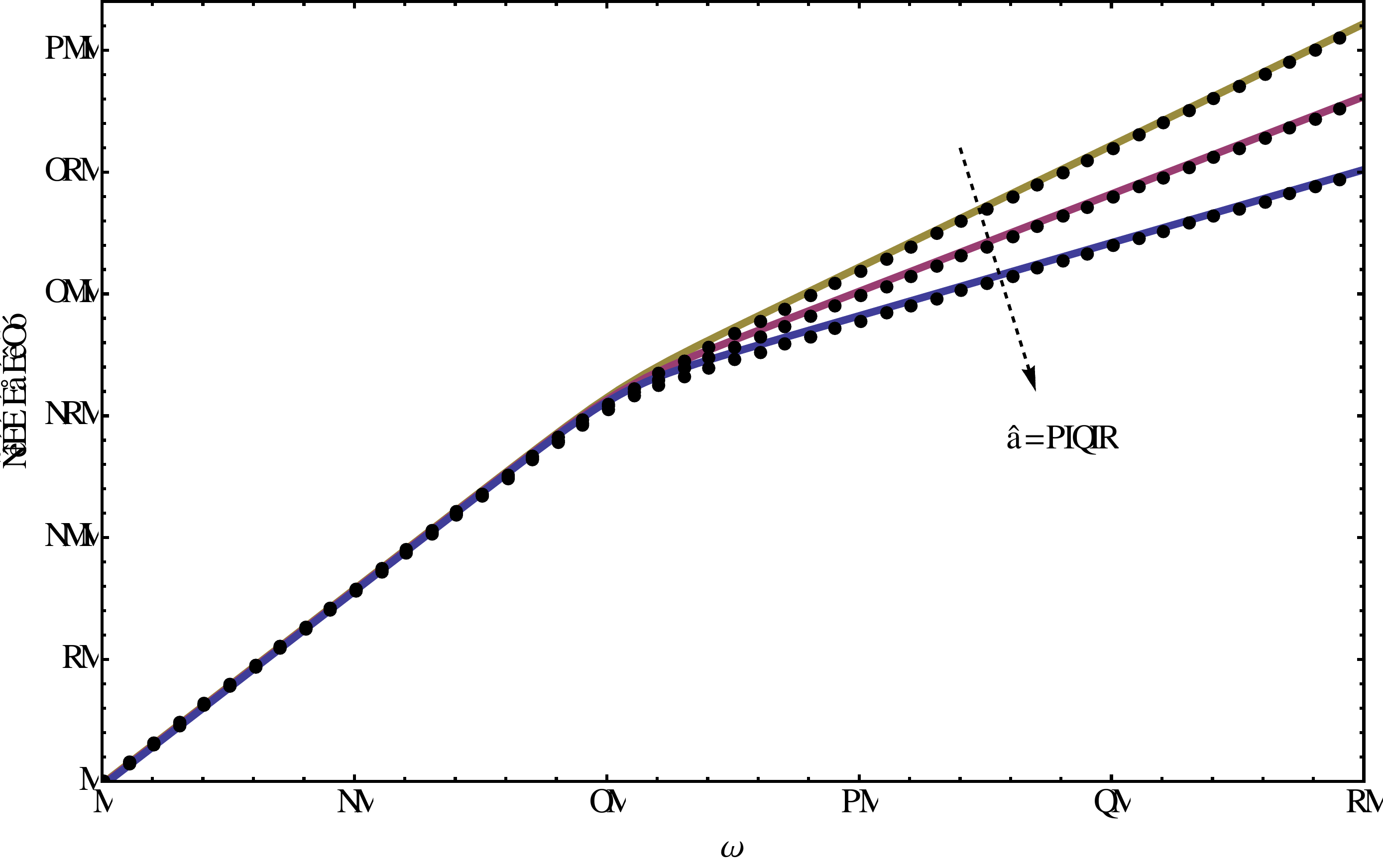}
\end{subfigure}
\caption{Free energy as a function of $\omega$, for $\Delta=20$, $\hbar=1$.
The top panel shows the case $k=1$ for different values of $F$;
the bottom panel the case $F=2$ for different values of $k$.
The solid lines show the analytical semiclassical approximation,
the dots are numerical data. 
The inset zooms in on the transition region, showing
that the semiclassical approximation deviates appreciably only in the
crossover regime, close to $\omega\sim\frac{\Delta}{\hbar}$.
}
\label{figure:semiclassical}
\end{figure}

\section{Numerical checks\label{section:numerical}}

In this section we are going to explore the model by means of numerical calculations.
In particular, we wish to check the validity of the semiclassical approximation for the partition function,
and study the dependence of the free energy on $\omega$ for general $F$ and $k$.
We perform a numerical evaluation of the partition function as
\be Z(F,k,\omega,\Delta,g,\hbar)=\textrm{Tr}e^{-H_n} \ee
where $H_n$ is the Hamiltonian restricted to the subspace with fixed total number of excitations $n$.

\begin{figure*}
\centering
\begin{subfigure}
  \centering
  \includegraphics[width=.45\linewidth]{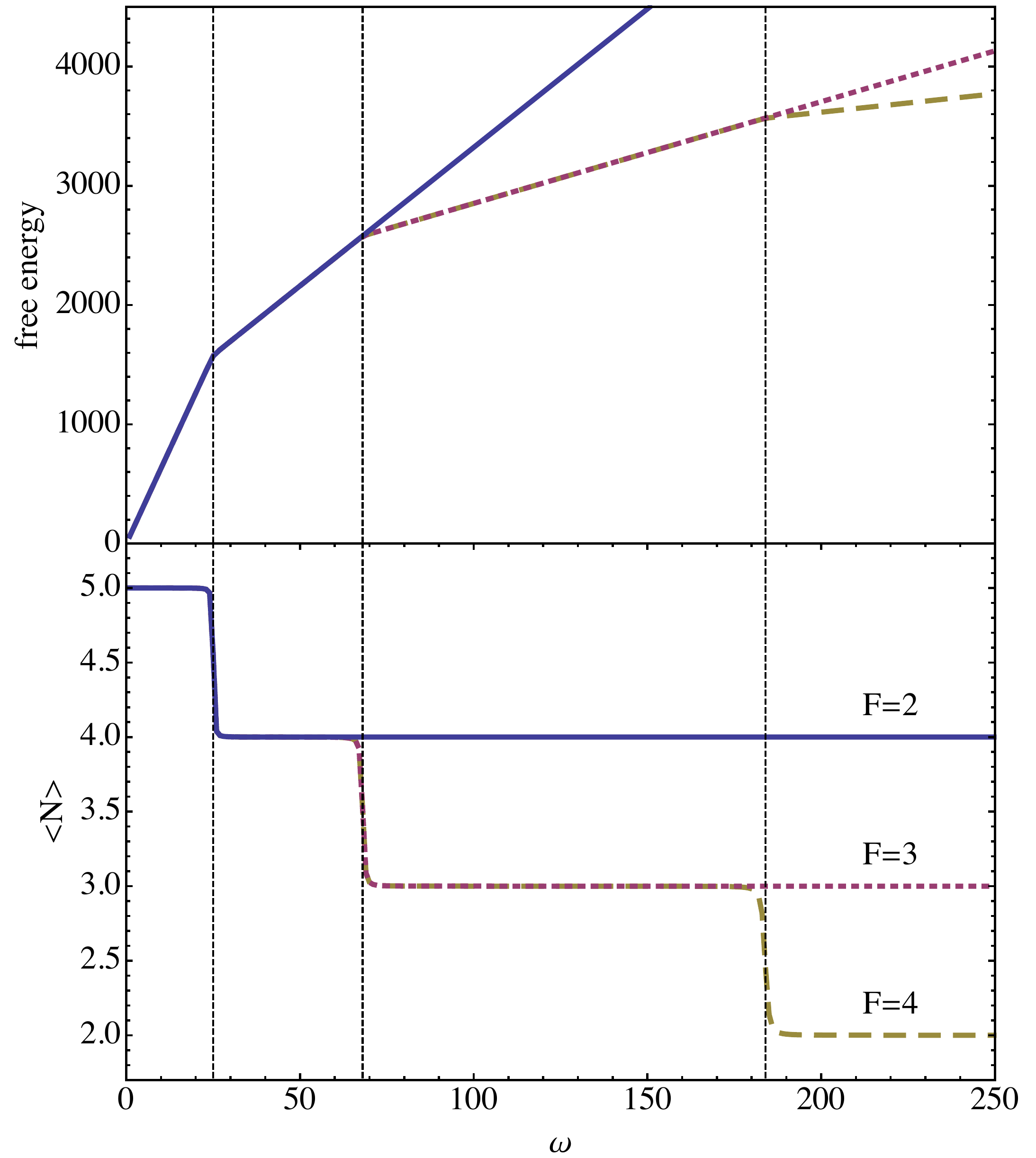}
\end{subfigure}%
\begin{subfigure}
  \centering
  \includegraphics[width=.45\linewidth]{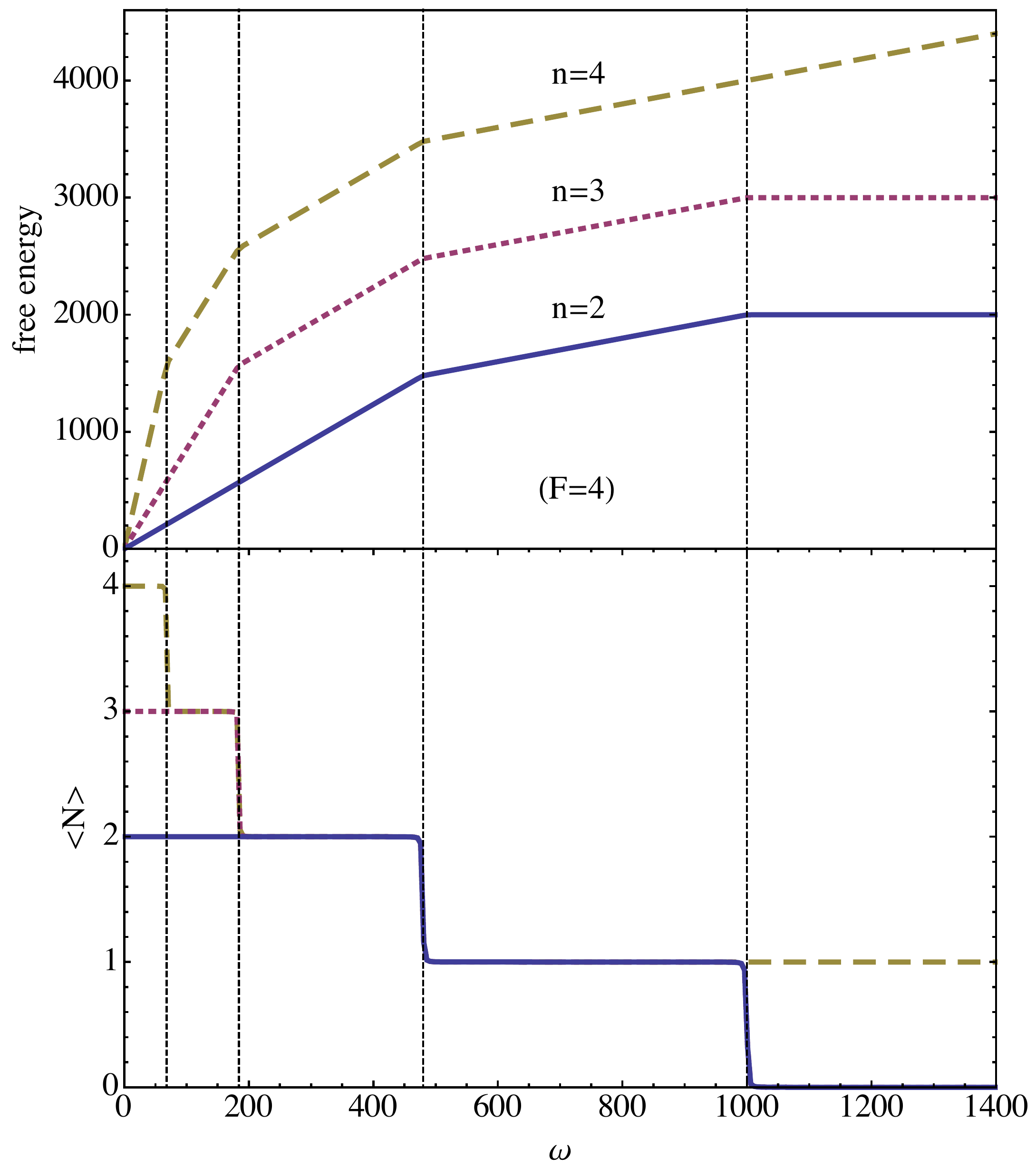}
\end{subfigure}
\caption{Free energy (top) and average of the number operator (bottom) as functions of $\omega$
for $q$-deformed harmonic oscillator, with $q=e^\hbar$, $\hbar=1$. The coupling is fixed to $\Delta=1000$, and $k=1$.
The left panel refers to the case with $n=5$ and shows different values of $F$.
The crossovers in the free energy correspond to crossovers in the number operator, although $\left<N\right>$
is not simply the derivative of the free energy with respect to $\omega$.
The right panel refers to the case with $F=4$ and shows different values of $n$.
When $n\leq (F-1)k$, $\left<N\right>$ is asymptotically zero, due to the saturation discussed in the text.
}
\label{figure:crossovers}
\end{figure*}

\begin{figure}
\centering
 \includegraphics[width=0.6\linewidth]{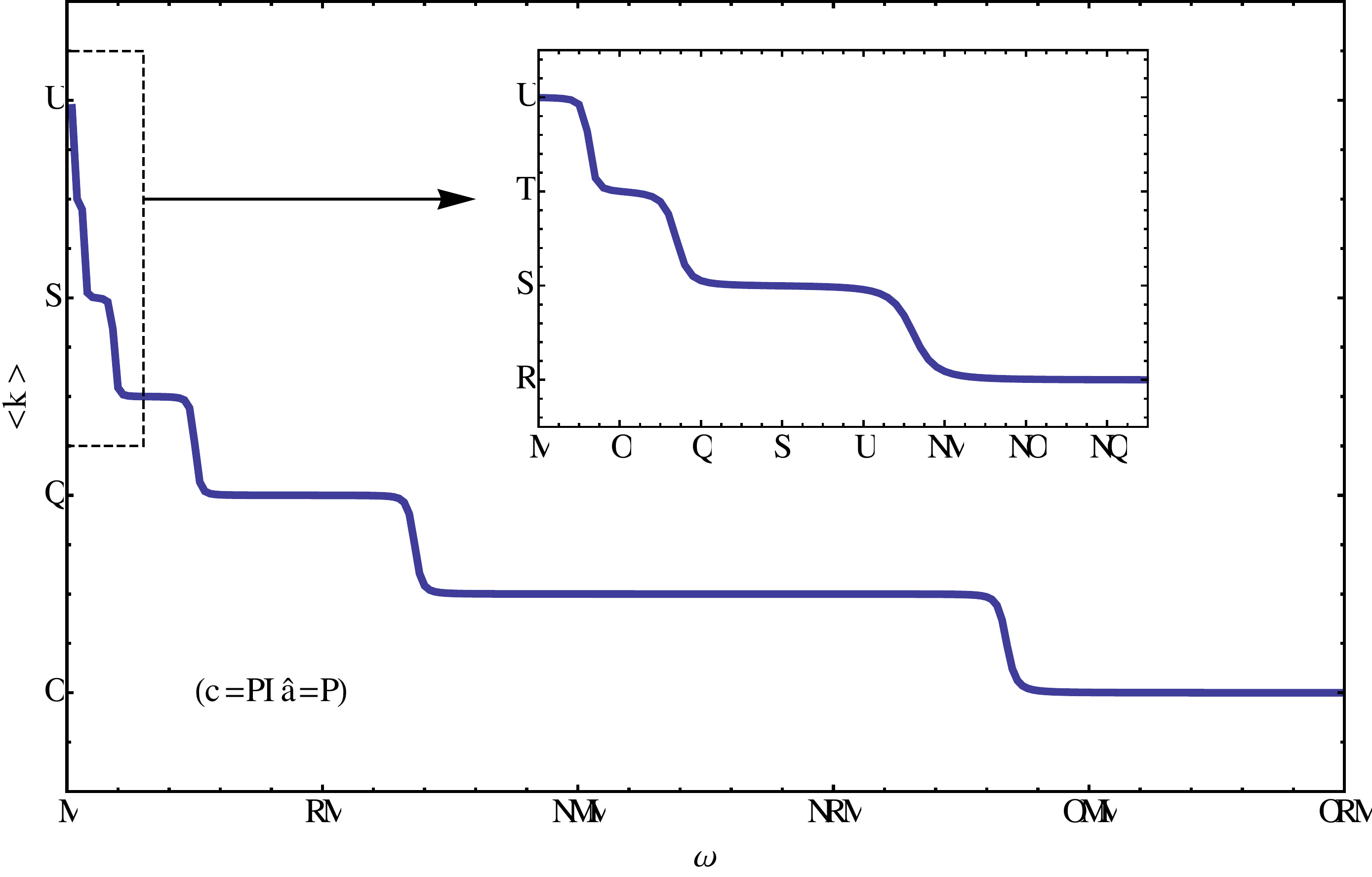}
\caption{Average of the bosonic number operator for $k=3$, $F=3$, $n=8$;
all other parameters are as in Fig.~\ref{figure:crossovers}.
The total number of crossovers is $k(F-1)$, corresponding to the non-saturated regime.
The inset zooms in close to $\omega=0$.
}
\label{figure:stairs}
\end{figure}

It turns out the the numerical diagonalization of the reduced Hamiltonians yields results in remarkable
agreement with the semiclassical approximation, even for $\hbar$ of order $1$ (Fig.~\ref{figure:semiclassical}).
As predicted by (\ref{eq:semiclassical_k}) and (\ref{eq:semiclassical_F}),
a crossover between two different regimes is present approximately at $\omega=\frac{\Delta}{\hbar}$.
The regime $\omega\leq\frac{\Delta}{\hbar}$ is superradiant, as signaled by the fact that
\begin{equation}
\big<a^{\dagger}a\big>= -\partial_{\omega}\log( Z(F,k,\hbar,\omega,\Delta,g))=\hbar(n+\delta- k (F-1))
\end{equation}
holds with $\delta=1$, whereas $\delta=0$ in the normal regime.
We remark that the semiclassical approximation was derived in the regime $n>k(F-1)$, where the dimension of the Hilbert space saturates to $F^k$ (and remains constant).
In the intermediate regime $n\leq k(F-1)$ one has
\be\big<a^{\dagger}a\big>= \hbar(n+\delta)  \ee
with $\delta$ having the same properties as above.

If we now turn on a deformation function
\be \Phi(x)=\frac{e^{\hbar x}-e^{-\hbar x}}{e^{\hbar}-e^{-\hbar}}  \ee
which corresponds to coupling parafermions to a $q-$deformed harmonic oscillator with $q=e^{\hbar}$, we observe a richer structure of crossovers.
Remark that by taking the derivative of the free energy with respect to $\omega$ one has direct access to the average $\big<\Phi(N)\big>$; however, we are interested in the average of the number operator $\big<N\big>$.
To this purpose, we perturb the original hamiltonian by adding a term of the form $\mu N$.
Computing the derivative of this modified free energy with respect to $\mu$,
and evaluating it at $\mu=0$, gives the expectation value of $N$ for the deformed oscillator.
We find that $\big<N\big>$ for generic $F$ presents multiple crossovers as a function of $\omega$
(Fig.~\ref{figure:crossovers}), resembling a staircase.
In the regime $n>k(F-1)$ we observe precisely $k(F-1)+1$ steps in the staircase,
starting from the value $n$ and going down to $n-k(F-1)$.
In the regime with $n\leq k(F-1)$ the staircase starts again from $n$ and goes down to $0$,
passing through all integer values in between (Fig,~\ref{figure:stairs}).
One expects that the $q$-deformed system converges to the undeformed case when $\hbar\to 0$;
this is indeed what we observe: regimes successively disappear as $\hbar$ is decreased,
until only two are left (not shown here).

Finally, we have performed numerical analysis also for the standard higher-spin JC model, 
by taking advantage of the connection described in Sec.~\ref{section:deformed},
and we observe that the phenomenology of the crossovers remains qualitatively the same as in the parafermionic case.
Multiple crossovers can be observed as well if the bosonic oscillator is deformed.

\section{Discussion}
In this work we have introduced a novel generalization of the Jaynes Cummings model by replacing spins with unconventional degrees of freedom represented by Fock parafermions and also by allowing the harmonic oscillator to be replaced by a generic deformed oscillator algebra. 
The coupling between parafermions and oscillator modes is linear;
this can be regarded as an approximation of the JC model describing the same type of phenomenology,
along the lines of existing literature on the JC-Hubbard model.
On the other hand, the unconventional form of the interaction permits to obtain some useful results.
We have solved exactly several instances of the model and we have derived 
a remarkably accurate semiclassical approximation for the canonical partition function of the model. 
The parafermionic model, like the ordinary JC model, exhibits a smooth crossover between a superradiant regime and a normal regime. 
In the case where the harmonic oscillator is replaced by a $q$-deformed oscillator (with real $q$) 
multiple crossovers emerge between different regimes with different average number of oscillator modes.

An interesting direction for future investigations regards the introduction
of more physical interaction terms.
Firstly, one can consider the ``four-legged vertex'' $a^\dagger\theta_i^\dagger \theta_i^2$,
in which interaction over a distance is mediated by the bosonic mode and the exchange of a parafermion.
This term does not break the global $U(1)$ symmetry associated with the conservation of $N_{tot}$ 
and it is nontrivial only in the truly parafermionic case $F>2$.
Secondly, one can introduce hopping terms for the parafermions 
and/or the deformed oscillator modes, thus formulating a sort of 
parafermionic Jaynes-Cummings-Hubbard model.
Finally, it would be interesting to explore the phase diagram of
a parafermionic Hubbard model without bosonic degrees of freedom.

\section{Acknowledgements}
We thank Giovanni Viola for useful advice and discussions.\\
Both authors acknowledge partial financial support from Fondo Sociale Europeo (Regione Lombardia), through the grant ``Dote ricerca''.

\end{document}